\documentclass[pra,showpacs,superscriptaddress,twocolumn,10pt]{revtex4-2}
\usepackage{braket}
\usepackage[cm]{fullpage}
\usepackage{graphicx}	
\usepackage[english]{babel}
\usepackage{amsmath}
\usepackage{amssymb}
\usepackage{cancel}
\usepackage{natbib}
\usepackage{comment}
\usepackage{color}
\usepackage[normalem]{ulem}
\usepackage{bbold}
\usepackage[outdir=./]{epstopdf}
\usepackage{here}

\newcommand{\tens} {\!\otimes\!}
\newcommand{\ie} {i.e.~}
\newcommand{\eg} {e.g.~}

\newcommand{\gbest}{\gamma_{\textrm{best}}}

\DeclareMathOperator{\tr}{tr}
\makeatletter
	\renewcommand{\maketag@@@}[1]{\hbox{\m@th\normalsize\normalfont#1}}
\makeatother
\begin{document}

\author{N.~Milazzo}
\affiliation{Institut f\"ur theoretische Physik, Universit\"at T\"{u}bingen, 72076 T\"ubingen, Germany}
\affiliation{LPTMS, CNRS, Univ.~Paris-Sud, Universit\'e Paris-Saclay, 91405 Orsay, France}
\author{D.~Braun}
\affiliation{Institut f\"ur theoretische Physik, Universit\"at T\"{u}bingen, 72076 T\"ubingen, Germany}
\author{O.~Giraud}
\affiliation{LPTMS, CNRS, Univ.~Paris-Sud, Universit\'e Paris-Saclay,
   91405 Orsay, France}

\begin{abstract}
With the advance of quantum information technology, the question of
how to most efficiently test quantum circuits is becoming of
increasing relevance. Here we introduce the statistics of lengths of
measurement sequences that allows one to certify entanglement across a
given bi-partition of a multi-qubit system over the possible sequence
of measurements of random unknown states, and identify the best
measurement strategies in the 
sense of the (on average) shortest measurement sequence of (multi-qubit)
Pauli-measurements.  The approach is based on the algorithm of
truncated moment sequences that allows one to deal naturally with
incomplete information, i.e.~information that does not fully specify
the quantum state. We find that the set of measurements corresponding
to diagonal matrix elements of the moment matrix of the state are
particularly efficient.  For symmetric states their number grows only
like the third power of the number $N$ of qubits.  Their efficiency
grows rapidly with $N$, leaving already for $N=4$ less than a fraction
$10^{-6}$ of 
randomly chosen entangled states undetected.
\end{abstract}


\title{Optimal measurement strategies for fast entanglement detection}

\date{February 13, 2019}

\maketitle

\section{Introduction}

With the availability of the first small quantum processors, the task
of characterizing such processors has become a key challenge.
Indeed, long before proving full functionality, one of the major
questions that faces a quantum processor 
 is whether it ``truly'' works quantum
mechanically --- or could rather be explained by classical
processes. Similar questions arise already at the level of a quantum
state: given a physical system in an unknown quantum state, can the
statistics arising from it be explained by a classical state? If
the state is fully characterized, one can apply non-classicality
measures to find out, but since a
mixed quantum state of $N$ qubits is specified by $d=2^{2N}-1$ real
parameters, it is clear that an answer based on full state quantum
tomography quickly becomes impractical.
In addition, one can only estimate expectation values based on 
averages over finitely many measurements that are themselves imperfect, and the 
resulting uncertainty can lead to nonphysical states in the 
inversion procedure underlying full quantum state tomography.  
More robust approaches to state tomography are maximum likelihood
estimation of the state
\cite{james_measurement_2001,blume-kohout_hedged_2010,hradil_quantum-state_1997,baumgratz_scalable_2013}
or Bayesian inference \cite{blume-kohout_optimal_2010,rau_evidence_2010},  
which output estimates of 
the state that are by construction {\em bona fide} physical states, as
well as ``self-consistent quantum-tomography'' not necessarily relying on
perfect measurements \cite{merkel_self-consistent_2013}, 
but none of these approaches remedies the efficiency problem.

Recent developments based on compressed sensing make use of prior
information of states. They provide a large gain in efficiency, in
particular for the typically low-rank states 
relevant for quantum information tasks
\cite{gross_quantum_2010,ohliger_continuous-variable_2011,ohliger_efficient_2013,kliesch_guaranteed_2017,flammia_quantum_2012},
or matrix-product states that describe
interacting condensed-matter systems in low dimensions
\cite{cramer_efficient_2010,lanyon_efficient_2016}. 
Machine learning  
aimed at determining by itself  what the best measurements are for a
certain task, or to recognize entanglement from measurement data,
was considered e.g.~in 
\cite{ferrie_self-guided_2014,wang_learning_2017,lu_separability-entanglement_2017,torlai_neural-network_2018},
but the efficiency of such approaches needs further study. Other proposals include few-copy multi-particle entanglement detection based on probabilistic verification \cite{saggio2018experimental,dimic2018single}.    

For testing quantum circuits, the
approach of randomized benchmarking has emerged
\cite{knill_randomized_2008,magesan_efficient_2012,gaebler_randomized_2012,chiribella_identification_2013}.   
Key to this approach 
is that for estimating fidelities between actual and ideal gate sets,
only low moments of the matrix elements are required.  In such a case,
averaging over the full unitary group can be replaced by averaging
over a unitary $t$-design \cite{dankert_exact_2009}, or producing
required input states by random
quantum circuits (see also \cite{Arnaud07}). 
\cite{schmiegelow_selective_2011,bendersky_selective_2009}
showed  
that a small number
of parameters of a quantum process can be efficiently obtained, but it
is not so clear what are the most relevant parameters that should be chosen. 

It is often stated that quantum states and quantum processors are much
harder to test and characterize than their classical analogues because
of the exponential growth of Hilbert space \cite{cramer_efficient_2010,lvovsky_continuous-variable_2009,titchener_scalable_2018}.  However, 
also classically the number of possible memory configurations of $N$
bits grows exponentially as $2^N$ --- and with $N$ of order
$10^{13}$ for a standard laptop computer, it is completely
out of the question to test all possible configurations. 
As costs for integration itself have decayed exponentially
according to Moore's law, for the same reason functional testing of classical
memory devices has evolved to
the most expensive (since time-consuming) part of the production of
integrated memory chips.  
 Functional
testing of classical memories has therefore evolved to {\em testing the
most critical known configurations with the goal of demonstrating failure of
memory cells as quickly as possible}.  ``Most critical'' depends on the
architecture of the chip, and information on its design goes
into the design of memory patterns to be tested.  For example, a cell
on a given bit-line might resist storing a ``0'' most likely if all
other cells on the same bit line contain a ``1''. In MRAM devices,
magnetic stray fields from a set of cells can destabilize others in
the vicinity when uniformly polarized, etc.

Quantum-information processing may still have a long way to go before such
economical pressure on functional testing will be felt.  At
the moment, rather than 
showing failure, one would like to prove basic quantum functionalities as
quickly as possible. Nevertheless, the principles of classical
functional testing can also provide guidance in the current state of
affairs in characterizing quantum processors and states: rather than aiming at
full quantum tomography, one may want to focus on producing
states that are likely to be particularly unstable, and show their
``functionality'' as 
quickly as possible.  In practice, this will need information
about the physical realization of the quantum processor, but in
the absence of such input, a reasonable target are highly entangled
states, or more generally, highly non-classical states known to be
prone to 
decoherence. Indeed, experimental efforts have concentrated early on
on producing such states (see
e.g.~\cite{haffner_scalable_2005,monz_14-qubit_2011,iskhakov_polarization-entangled_2012,mcconnell_entanglement_2015,bohnet_quantum_2016}
for states with large numbers of entangled particles).\\

The question 
arises then: what is the most efficient 
measurement strategy to prove that such a state is entangled (or more
generally: non-classical)? I.e.~what would you
choose to measure first, second, and so on, in order to be able to
prove as quickly as possible with the limited knowledge
about the  
state that you will gain 
from those measurements, chosen from a given set, that the state is
entangled? What are the 
minimum and average numbers of measurements needed to prove
entanglement, or more generally the statistics of the length of
measurement sequences when going down a certain path of measurements? 

These are the questions that we start to answer in the present
paper. Note that 
this is not about choosing optimal entanglement witnesses, 
but rather about deciding whether the intersection of hyperplanes
defined by the  
expectation values of certain observables cuts the set of separable
states or not (see Fig.~\ref{sketch}).  Perfectly suited for answering these
questions is the 
formalism of ``truncated moment sequences'' (TMS) that we introduced
in \cite{tms} to the analysis of entanglement.  The TMS problem aims
at finding a probability measure for which only some moments are
known. If the probability measure is furthermore constrained to be
supported on a compact set $K$, the problem is known as the $K$-TMS
problem. As will be reviewed below, it can be solved with a hierarchy
of flat extensions that 
maps onto a convex optimization algorithm, using a semidefinite
relaxation procedure. 
Each expectation value
can be associated with a moment of a measure, and instead of fixing
all moments up to a certain order as in the standard TMS algorithm,
one might just specify any set  $\mathcal{A}$ of 
moments.  The problem of deciding whether a classical measure that reproduces
all these moments exists is then known as the
''$\mathcal{A}K$-TMS problem'' \cite{ak}. It can still be solved  with
a convex optimization algorithm.

In the present work we exploit this approach in order to obtain
the statistics of lengths of measurement sequences in the
simplest case of two qubits depending on the chosen measurement
strategy.  For larger numbers of qubits, the 
full numerical solution of the  $\mathcal{A}K$-TMS problem becomes too
demanding, but it turns out that surprisingly efficient sufficient conditions
for entanglement can be obtained for symmetric states from the diagonal matrix elements of
the moment matrix  used in the
approach  (see below for a definition). These correspond to certain
linear combinations of  
expectation values of (possibly multipartite) measurements in the
Pauli basis and have to 
be positive for a solution of the $\mathcal{A}K$-TMS problem to
exist. Checking the positivity of moment matrices 
is in fact the
first step in the TMS-algorithm, and negativity of any of the diagonal
matrix elements hence witnesses entanglement.   With these we can 
find numerical estimates of the fraction of randomly drawn states that
are already 
detected as entangled by just measuring the observables corresponding
to the diagonal matrix elements of the moment  matrix.

Similar ideas for certifying entanglement with incomplete measurements
were already 
considered in \cite{mari_directly_2011} for continuous variable
systems. Here we focus on the statistics of
lengths of sequences of measurements for multi-qubit systems and the
insights that can be 
drawn from the TMS algorithm which we review in the next section,
before applying it to incomplete measurements.

\section{Framework and notation}
We now briefly summarize the TMS algorithm approach described in
detail in \cite{tms}, which will be the framework for the following
sections. The basic idea is to map the quantum entanglement problem
onto the mathematically well-studied truncated moment problem. Indeed,
finding whether an arbitrary multipartite state can be decomposed into
product states corresponds to finding about the existence of a
probability distribution whose lowest-order moments are
fixed. Analytically, the mapping allows one to make use of theorems from
the TMS  literature providing necessary and
sufficient separability conditions; numerically, semidefinite
optimization techniques yield an algorithm which gives a certificate
of entanglement or separability. The algorithm applies --- at least in
principle --- to arbitrary 
quantum states with arbitrary number of constituents and arbitrary
symmetries between the subparts. The general case is dealt with in
\cite{tms}; we only recall here the main key points for the case of
symmetric states of qubits, defined as mixtures of symmetric pure
states (the latter are invariant under any permutation of the
qubits). To do so, we will use a convenient representation in terms of
symmetric tensors which was introduced in \cite{tens}, generalizing
the Bloch sphere picture of spins$-1/2$. We can write a generic state
$ \rho $ of a spin-$j$ state as 
\begin{equation}
\rho=\frac{1}{2^N} \sum_{\mu_1,\mu_2,...,\mu_N=0}^3X_{\mu_1\mu_2...\mu_N} P_s( \sigma_{\mu_1} \tens ... \tens \sigma_{\mu_N} )P^{\dagger}_s 
\end{equation}
where $ \sigma_0 $ is the $ 2\times 2 $ identity matrix, $ \sigma_1,
\sigma_2, \sigma_3 $ are the Pauli matrices and $ P_s $ is the
projector onto the symmetric subspace spanned by the Dicke states
$\ket{j,m}$ (eigenstates of pseudo-angular momentum 
component $J_z$ and with total angular momentum quantum number $j$).  
They
can also be seen as symmetric states of $N=2j$ spins-1/2 (or qubits). 
The tensor $ X_{\mu_1\mu_2...\mu_N} $ is then given by
\begin{equation}
X_{\mu_1\mu_2...\mu_N}=\tr(\rho  \sigma_{\mu_1} \tens ... \tens \sigma_{\mu_N} ),
\label{tensrepr}
\end{equation}
with $0 \leq \mu_i\leq 3$. It is real and invariant under permutation of indices, and verifies $X_{00...0} = \tr(\rho) = 1$. Moreover, it has the property that 
\begin{equation}
\sum_{a=1}^3X_{aa\mu_3\mu_4...\mu_N}=X_{00\mu_3\mu_4...\mu_N}
\label{traceless_general}
\end{equation}
for any choice of the $\mu_i$.
A separable pure state can be seen as a spin-coherent  state, which in
the representation \eqref{tensrepr} has tensor entries
$X_{\mu_1\mu_2...\mu_N}=n_{\mu_1}n_{\mu_2}...n_{\mu_n}$, with $n_0=1$
and $(n_1,n_2,n_3)$ the unit vector giving the direction of the
coherent state on the Bloch sphere. In terms of this tensor
representation,  a symmetric state is separable if and only if its tensor representation can be written as
\begin{equation}
X_{\mu_1\mu_2...\mu_N}=\sum_i \omega_i n^{(i)}_{\mu_1}n^{(i)}_{\mu_2}...n^{(i)}_{\mu_N},\qquad \omega_i\geq 0,
\label{sep}
\end{equation}
where $n_0^{(i)} = 1$ and $\mathbf{n}^{(i)}$ is the Bloch vector of the single qubit. If we express \eqref{sep} in the equivalent integral form 
\begin{equation}
X_{\mu_1\mu_2...\mu_N}=\int_K x_{\mu_1}x_{\mu_2}..x_{\mu_N} d\mu(\mathbf{x})
\label{map}
\end{equation}\\
with $ x_0=1 $, $ K=\lbrace \mathbf{x}\in \mathbb{R}^3: x_1^2+x_2^2+x_3^2=1 \rbrace $ the unit sphere and $ d\mu(\mathbf{x})=\sum_i \omega_i\delta(\mathbf{x}-\mathbf{n}^{(i)}) $ a positive measure on $K$, we can say that a symmetric state is separable if and only if there exists a positive measure $ d\mu $ supported by $K$ such that all entries of the tensor $X_{\mu_1\mu_2...\mu_N}$ are given by moments of that measure. 

Problems of this type are known as $K$-TMS problems, or
$\mathcal{A}K$-TMS problems in the case of  partial knowledge of a
state where only a subset of the moments, specified by the set
$\mathcal{A}$, is known. They can be solved by a semidefinite
relaxation procedure. The algorithm proposed in \cite{tms} uses indeed
semidefinite programming (SDP) and the concept of "extensions",
already introduced in \cite{ext}, but based on a matrix of moments and
a theorem in the theory of moment sequences. In order to present more
clearly the mathematical setting for the $\mathcal{A}K$-TMS problem,
we introduce a more compact notation for Eq.~\eqref{map}. For any
$N$-tuple $(\mu_1, . . . ,\mu_N)$ we define a triplet $\alpha =
(\alpha_1,\alpha_2,\alpha_3)$ of integers such that
$x_{\mu_1}x_{\mu_2} ... x_{\mu_N} = x^{\alpha}$, where we use the
notation $x^{\alpha} = x_1^{\alpha_1} x_2^{\alpha_2}
x_3^{\alpha_3}$. The degree of the monomial $ x^{\alpha} $ is $\vert
\alpha\vert=\sum_i \alpha_i$. We then set  $y_\alpha\equiv
X_{\mu_1\mu_2...\mu_N}$. The $(y_\alpha)_{|\alpha|\leq d}$ is a TMS,
that is, a sequence of moments of $\mu$ truncated at degree $d$. In
case only a subset $\alpha\in\mathcal{A}$ of these moments is known,
we consider the TMS $(y_\alpha)_{\alpha\in\mathcal{A}}$. With this
notation we can rewrite \eqref{map} as  
\begin{equation}
y_{\alpha}=\int_K x^{\alpha} d\mu(\mathbf{x}).
\label{map2}
\end{equation}
To a TMS $y$ of degree \textit{d}, for any integer $k\leq d/2$, we can
associate a matrix $M_k(y)$ defined by $M_k(y)_{\alpha
  \beta}=y_{\alpha+\beta}$ with $ \vert \alpha\vert,\vert
\beta\vert\leq k $, which we call the $k$th order moment matrix. A
necessary condition for a TMS to admit a representing measure is that
the moment matrix of any order be positive semidefinite. A second
necessary condition can be obtained from the polynomial constraint
$x_1^2+x_2^2+x_3^2=1$ which defines the set $K$. For even degree
\textit{d} we define a "shifted TMS" of degree $d-2$, and its
moment matrix of order $ k-1 $ is called the \textit{k}th-order
localizing matrix of $y$. It
is necessarily positive semidefinite if a TMS admits a representing
measure.

Beyond these two necessary conditions, a sufficient condition was obtained in \cite{rank} for even-degree TMS. Namely, if a TMS \textit{z} of even degree $2k$ is such that 
\begin{equation}
\textrm{rank} M_k(z)=\textrm{rank} M_{k-1}(z),
\label{rank}
\end{equation}
then the TMS \textit{z} admits a representing measure. As the above condition is only sufficient, a TMS admitting a representing measure does not necessarily fulfill it, but one can always search for an extension of it which does. An extension of a TMS \textit{y} of degree \textit{d} is defined as any TMS \textit{z} of degree \textit{2k} with $2k > d$, such that $z_{\alpha} = y_{\alpha}$ for all $\alpha\in\mathcal{A}$. An extension \textit{z} is called flat if it satisfies Eq. \eqref{rank}. If \textit{z} verifies the sufficient conditions above, then it has a representing measure, and so does \textit{y} as a restriction of \textit{z}. Then it is possible to formulate a necessary and sufficient condition for the existence of a representing measure as follows:\\
\textit{Theorem} A state $ \rho $ is separable if and only if its
coordinates $ X_{\mu_1\mu_2...\mu_N} $ are mapped to a TMS
$(y_{\alpha})_{\alpha\in\mathcal{A}} $ such that there exists a flat
extension $(z_{\beta})_{\vert \beta \vert\leq 2k}$ with $2k>d$, and
whose corresponding $k$th order  moment and localizing matrices are
positive semidefinite.

This necessary and sufficient condition can be translated into an
algorithm looking for flat extensions of the TMS $y$ associated with a quantum state $\rho$. One runs the algorithm with input the state $ \rho $ (that means fixing $y_{\alpha}$ for all $\alpha\in\mathcal{A}$), starting from the lowest possible extension order $k$. If the corresponding SDP is "infeasible", then the conditions of the theorem are not satisfied and the TMS admits no representing measure $ d\mu$, which means that the quantum state whose coordinates are given by $ y_{\alpha} $ is entangled. If, on the contrary, the SDP problem is "feasible", then the TMS admits a representing measure, and the corresponding quantum state is separable. The algorithm also extends to the case of non-symmetric states (see \cite{tms} for further detail).

\section{Unordered measurements}\label{sec.unordered}

\subsection{Goal}

Let us now consider the question raised in the introduction.
Our goal is to identify the smallest set of measurements that
should be performed on an unknown spin state to detect that it is
entangled.  This is possible in a real experiment when many identical
copies of the same state are available, so that a different 
measurement can be performed on each copy. We will first discuss the case of symmetric two-qubit states, which, as we will see in detail, already presents some complexity.
In this case the PPT criterion \cite{PeresPT,HorPT} applied to a partially known density matrix would also provide a way of detecting entanglement via SDP. Nevertheless, we will use our TMS approach, since it allows for a straightforward generalization to arbitrary number of qubits and moreover it applies SDP to the matrix of moments, whose entries are directly given by measurement results.

\subsection{Symmetries and measurements}
\label{symandmeas}

For a symmetric two-qubit state $ \rho $, Eq.~\eqref{tensrepr} with $N=2$ gives
\begin{equation}
X_{\mu_1\mu_2}=tr(\rho \sigma_{\mu_1} \otimes \sigma_{\mu_2}  )
\end{equation}
with $0 \leq \mu_i\leq 3$ and $
(\sigma_0,\sigma_1,\sigma_2,\sigma_3)\equiv
(\mathbb{1},\sigma_x,\sigma_y,\sigma_z) $. In this case the tensor $
X_{\mu_1\mu_2} $ reduces to a $ 4\times 4 $ real symmetric matrix. Its
10 entries $X_{\mu_1\mu_2}$ with $\mu_1\leq\mu_2$ can be seen as the
result of the measurement of the joint operator $ \sigma_{\mu_1}
\otimes \sigma_{\mu_2}$. We now ask which are the possible
measurements that we can perform 
and how many there are; 
 the observables considered are the most simple, \ie Pauli spin
 operators.  Let us denote these inequivalent measurement operators as 
\begin{equation}
\mathcal{M}=\lbrace M_x,M_y,M_z,M_{xx},M_{xy},M_{xz},M_{yy},M_{yz},M_{zz}  \rbrace
\label{setM}
\end{equation}
(we omit the identity operator corresponding to $ X_{00}=1$, and we always order sets of measurements in degree-lexicographic order).
For instance, $ M_x$ is the measurement of $ \mathbb{1} $ on the first
qubit and of  $ \sigma_{x} $ on the second one (or the reverse), while
$ M_{xx}$ is the measurement of the joint operator
$\sigma_{x}\otimes\sigma_{x}$. Since the tensor $ X_{\mu_1\mu_2} $ is
such that 
\begin{equation}
\label{traceless}
\sum_{i=1}^3 X_{ii}=X_{00},
\end{equation}
only two of the three diagonal entries are independent, and measuring two out of three of the observables $M_{xx}$, $M_{yy}$ and $M_{zz}$ yields the third value. Thus, carrying a tomography to its end for a single spin-1 state consists in measuring 8 observables in total. 

Our aim is to find the probability that a state is detected as entangled if only the result of measurements of a certain subset of these 8 observables is known. Let us first of all observe that these probabilities should not depend on the choice of the reference frame for the axes along which the measurement is performed. As a consequence, the results for equivalent measurements in different directions should be the same. We will therefore only consider sets which are non-equivalent under permutation of the axes, that is, sets that are unchanged under transpositions $ \lbrace P_{xy}, P_{xz}, P_{yz}\rbrace $, which exchange two axes, and cyclic permutations $P_{yzx}$ and $P_{zxy}$. 

We will consider all possible non-equivalent sets of $k$ measurements,
with $ 1 \leq k \leq 8 $, disregarding the order 
of measurements within a set.
For sets of length $ k=1 $ we can easily see that the non-equivalent
measurements are only three: $ M_x, M_{xx}, M_{xy} $. {Indeed, the
  local measurements $M_x,M_y,M_z $ are equivalent, as well as the
  two-qubit ''diagonal'' measurements $ M_{xx},M_{yy},M_{zz}$ (giving the
  diagonal entries of matrix $(X_{\mu\nu})_{1\leq \mu,\nu\leq 3}$),
  and also the two-qubit ''off-diagonal'' measurements  $
  M_{xy},M_{xz},M_{yz} $ (giving its off-diagonal entries). }For $
k=2$, there are $28$ possible pairs, among which only 9 are
inequivalent, namely $\lbrace M_x,M_y\rbrace,\lbrace
M_x,M_{xx}\rbrace,\lbrace M_x,M_{xy}\rbrace,\lbrace
M_x,M_{yy}\rbrace,\lbrace M_x,M_{yz}\rbrace,\\\lbrace
M_{xx},M_{xy}\rbrace,\lbrace M_{xx},M_{yy}\rbrace,\lbrace
M_{xx},M_{yz}\rbrace,\lbrace M_{xy},M_{xz}\rbrace$. We denote by $m_k$
the number of non-equivalent sets of $k$ measurements, and we report it
in Table \ref{t1}. The corresponding complete lists of measurements
for all $ k $ is given in Appendix \ref{app}. 
 \begin{center}
 \begin{table}
  \begin{tabular}{ c|c|c|c|c|c|c|c|c }
    \hline
    k & 1 & 2 & 3 & 4 & 5 & 6 & 7 & 8 \\ \hline
    $m_k$ (unordered) & 3 & 9 & 19 & 26 & 23 & 14 & 5 & 1\\ \hline
        $m_k'$ (ordered; $ M_{xx} $ fixed) & 1 & 5 & 26 & 128 & 524 & 1604 & 3228 & 3228 \\
       \hline
  \end{tabular}
  \caption{First line: Number $ m_k $ of non-equivalent unordered sets of measurements for $ 1\leq k \leq 8 $. Second line: Number $ m_k$ of non-equivalent ordered sequences of measurements for $ 1\leq k \leq 8 $.} 
  \label{t1}
 \end{table}
 \end{center}

For each $k$, our question reduces to finding out which set of measurements, among the $m_k$ possible ones, yields the highest entanglement detection probability. Note that performing $k$ measurements is not exactly equivalent to having $k$ fixed moments. Indeed, since moments are related by Eq.~\eqref{traceless}, measuring $M_{xx}$ and $M_{yy}$ fixes {the three moments $X_{11}, X_{22}$ and $X_{33}$}. Any measurement set of length $k$ containing both $M_{xx}$ and $M_{yy}$ will in fact correspond to a TMS with $k+1$ moments fixed. We therefore always discard $M_{zz}$ from the measurement sets.

\subsection{Set probabilities}
In terms of the TMS algorithm, performing a measurement means obtaining a value of a tensor entry $X_{\mu_1\mu_2...\mu_N}$, or equivalently of a moment $y_{\alpha}$. Performing $k$ measurements means that the $k$ moments $y_{\alpha}$ corresponding to those measurements are fixed, as well as all moments obtained via relation \eqref{traceless_general}.

For a given number $k$ of measurements, we indicate a specific set of measurements among the $m_k$ possible ones as $ \lbrace M\rbrace_I $. For instance, if $ k=3 $, we could have $ I=\lbrace x,y,zz \rbrace $ which corresponds to the set of measurements $ \lbrace M_x,M_y,M_{zz}\rbrace $.

If we consider a fixed $k$ and a fixed subset $ \lbrace M\rbrace_I $ of the set of observables $ \mathcal{M} $, we denote the sample space of outcomes of the $\mathcal{A}K$-TMS algorithm applied to the moments $(y_{\alpha})_{\alpha\in\mathcal{A}}$ of an entangled state
 as $\Omega_I$. It contains two possible outcomes, to which a probability can be assigned: detecting the state as entangled (if the associated SDP is infeasible, \ie if the state is entangled), with probability $P(E, \lbrace M\rbrace_I)$, or \textit{not} detecting it as entangled (if the SDP is feasible, \ie if the state with such moments fixed is still compatible with a separable state), with probability $P(\bar{E}, \lbrace M\rbrace_I)$. To shorten notation we may denote $P(E, \lbrace M\rbrace_I)$ as $p_I^{(k)}$, which entails $P(\bar{E}, \lbrace M\rbrace_I)=1-p_I^{(k)}$.
 
\begin{figure}[htpb]
\includegraphics[scale=0.72]{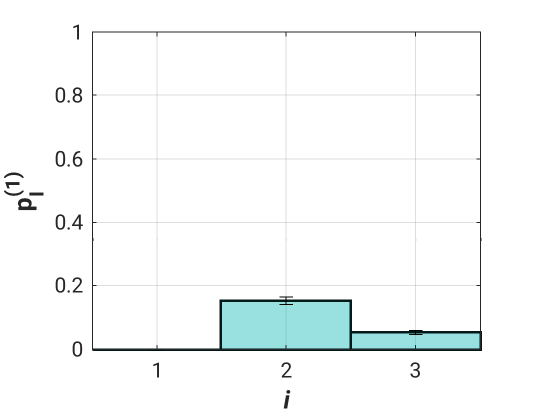}
\includegraphics[scale=0.72]{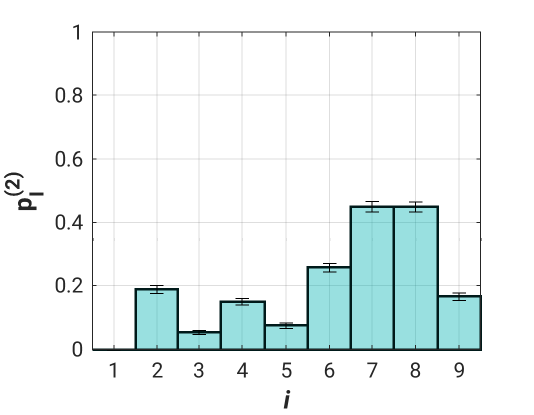}
\includegraphics[scale=0.72]{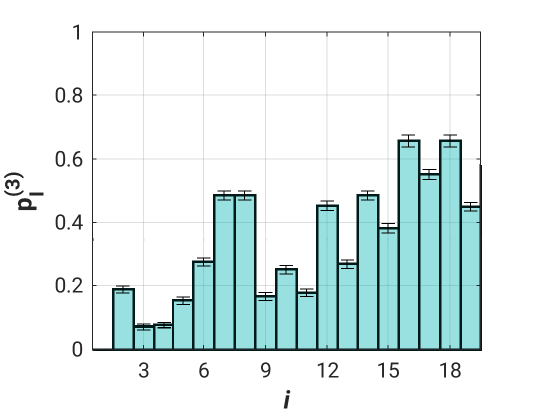}
\includegraphics[scale=0.72]{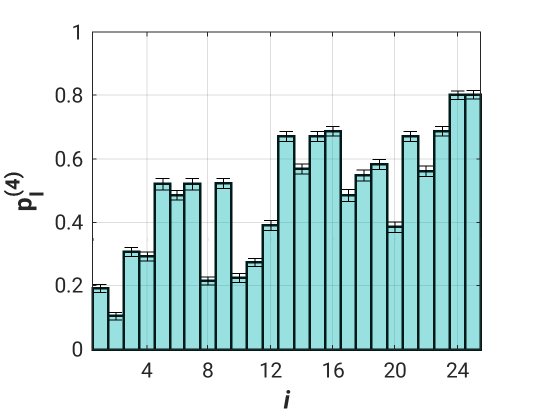}
\caption{{Probabilities $p^{(k)}_I$ 
of detecting entanglement in a symmetric state of two qubits with measurement set $ I $ of cardinality $k$, $ 1\leq k\leq 4 $, as a function of the label $i$ of the set $I$ ($1\leq i\leq m_k$). The associated error bars are given by the difference between the maximum and the minimum of the fluctuations observed for 1000 different samples of size $ 4\times 10^4 $ randomly extracted from the initial sample considered. The set of measurements $\{M\}_I$ corresponding to each label is given in Appendix \ref{app}}.}
\label{prob}
\end{figure}

\begin{figure}[htpb]
\includegraphics[scale=0.8]{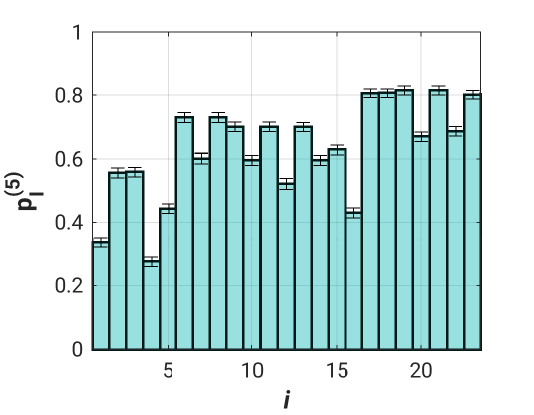}
\includegraphics[scale=0.8]{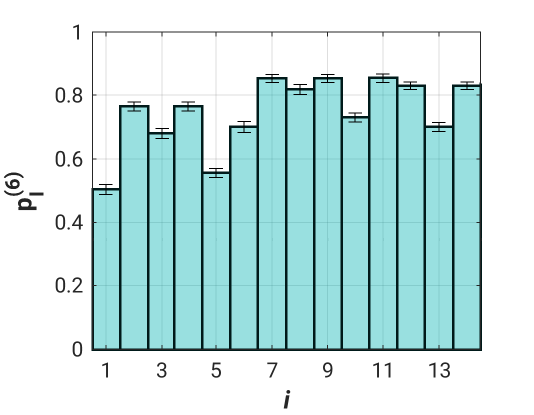}
\includegraphics[scale=0.8]{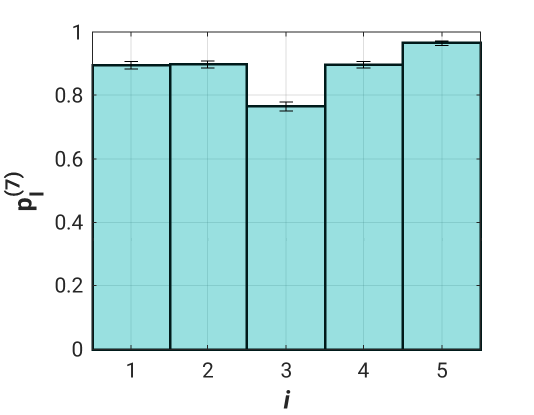}
\caption{Probabilities $p^{(k)}_I$ for $ 5\leq k\leq 7$, same as Fig.~\ref{prob}. }
\label{prob1}
\end{figure}

These probabilities can be estimated by running the TMS algorithm for
each $k$ and each $I$, testing all the $m_k$ possible sets of
measurements. Note that $ p_I^{(k)} $ always increases, in the sense
that $p_J^{(k')}\leq p_I^{(k)}$ for $J\subset I$. Indeed, the
probability not to detect entanglement with more and more measurements
goes down with the number of measurements. In other words, fixing more
moments $y_\alpha$ reduces the probability of finding a measure $\mu$
with such moments. Once all eight measurements are done the state is
fixed uniquely, so that for entangled states $p_I^{(8)}=1$.  To
estimate the values for the probabilities $ p_I^{(k)} $, we  sample
states from the set of symmetric two-qubit states. We generated
$5\times 10^4$ random states drawn from the Hilbert-Schmidt ensemble
of matrices $ \rho=\frac{GG^{\dagger}}{tr(GG^{\dagger})} $, with $G$ a
complex matrix with independent Gaussian entries (following
\cite{rand}).  
Among them were 1843 separable states that we discarded, implying
the normalization condition $p_I^{(8)}=1$ for full tomography. 
For each measurement set $ \lbrace M\rbrace_I $ and each entangled state in our sample the TMS algorithm was run with the corresponding moments fixed; the results for the probabilities $ p_I^{(k)} $  are reported in Figs.~\ref{prob} and \ref{prob1}.

{Some probabilities appear to be equal. This is for instance the
  case of probabilities labeled 16 and 18 for $k=3$. This is a
  consequence of an additional symmetry due to the linear equations
  that measurement results must satisfy. In the case $k=3$, labels 16
  and 18 correspond to $\lbrace M_{xx},M_{xy},M_{yy} \rbrace $ and
  $\lbrace M_{xx},M_{xz},M_{yy} \rbrace$ respectively. Since, as we
  already mentioned, knowing the result of any two diagonal
  measurements gives the third one because of Eq.~\eqref{traceless},
  the information acquired by measuring the observables corresponding
  to labels 16 and 18 is equivalent, and therefore 
  the probabilities must be equal.

The optimal choice of measurements $\{M\}_{I_{\textrm{opt}}}$ at fixed $k$
corresponds to the sets giving the highest probability of detecting
entanglement. For $k=1$ the highest value of $p^{(1)}_I$ corresponds
to the measurement 
\#2, $\{M_{xx}\}$. For $ k=2$ it corresponds to \#7,
$\{M_{xx},M_{yy}\}$. For $ k=3 $ the highest values correspond to two
measurements: \#16, $\lbrace M_{xx},M_{xy},M_{yy} \rbrace $, and \#18,
$\lbrace M_{xx},M_{xz},M_{yy} \rbrace$. For $k=4$ it corresponds to
\#23, $\lbrace M_{xx}, M_{xy}, M_{xz}, M_{yy} \rbrace $, \#24,
$\lbrace M_{xx}, M_{xy}, M_{xz}, M_{yz} \rbrace $, and \#25, $\lbrace
M_{xx}, M_{xz}, M_{yy}, M_{yz} \rbrace$. Again, the degeneracy of the
optimal set reflects the equivalence of the corresponding sets once
\eqref{traceless} is taken into account. For $k\geq 2$, the sets
$\{M\}_{I_{\textrm{opt}}}$ in fact correspond to cases where measuring
two observables fixes three moments.

\subsection{Quantumness}
For a fixed set of measurements $ M_I $ one can ask whether the rate
of detected entangled states depends on \textit{how quantum} a state
is. For an arbitrary state $ \rho $,  quantumness may be defined in
several different ways; we follow here the definition given in
\cite{quant}, based on spin-coherent states. These are a
generalization of the usual coherent states of the  
harmonic oscillator used in quantum optics to spins; 
they correspond to spin states which minimize a particular uncertainty relation, and they move as classical phase space points under a Hamiltonian linear in the angular momentum operators \cite{CohstDicke, spincohst}. As any spin-1/2 pure state $\ket{\phi}$ has this property, an arbitrary $N$-qubit spin-coherent state can be defined as $\ket{\phi}^{\otimes N}$ with $\ket{\phi}$ a one-qubit state.

Quantumness is then defined as the Hilbert-Schmidt
distance to the convex set $ \mathcal{C} $ of classical spin states
\cite{class}, that is the ensemble of all density matrices which can
be expressed as a mixture of spin-coherent states with positive
weights (or in other words the set $ \mathcal{C} $ is the convex hull
of spin-coherent states). Namely, the quantumness $ Q(\rho) $ is given
by 
\begin{equation}
Q(\rho)=\underset{\rho_c\in\mathcal{C}}{\min}\Vert\rho-\rho_c\Vert
\end{equation}
where $ \Vert O\Vert=\sqrt{Tr(O^{\dagger}O)} $ is the Hilbert-Schmidt
norm. For all $ \rho $ the property $ Q(\rho)\geq 0 $ holds, with
equality for classical states $ \rho\in\mathcal{C} $. Results are
reported in Fig.~\ref{q}, up to $ k=4 $ for the optimal sets of
measurements $\{M\}_{I_{\textrm{opt}}}$ given above. We can observe
that the rate of detected entangled states increases with the
quantumness of the states, or in other words, the more quantum a
state is the faster it is detected as entangled.
\begin{figure}
\includegraphics[scale=0.9]{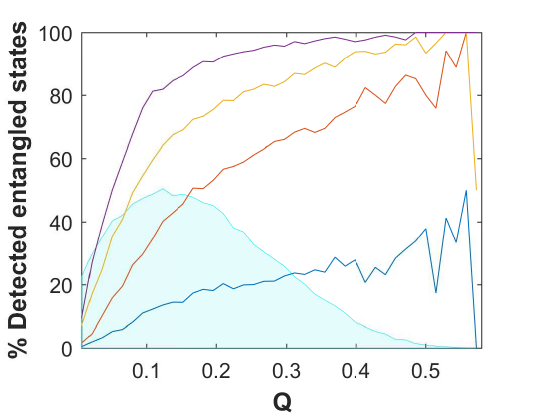}
\caption{
Percentage of detected entangled states for the optimal sets of
  measurements $ \lbrace M\rbrace_I $ for $k=1$ to $4$ (solid lines
  from bottom to top), as a function of quantumness 
for symmetric states of two qubits. 
The shaded area in the background represents the distribution of quantumness $ Q $ (bin width $ 0.015 $) of the total number of states (multiplied by a factor $ 2\cdot 10^{-2} $); the first bin contains entangled states with $ Q $ between $ 10^{-4} $ and $ 10^{-2} $. The distribution shows that there are very few states for the highest values of quantumness, which explains the large statistical errors at maximum quantumness.}
\label{q}
\end{figure}

\section{Ordered measurements}

\subsection{The setting}
{

In the previous section we assumed that $k$ observables are measured  among the 8 possible ones and that the TMS algorithm is subsequently run. Of course, we can imagine a different experimental protocol where we would perform a measurement, run the TMS algorithm with a single moment fixed, and then, only in the case where the state is not detected as entangled, perform a second measurement and run the TMS algorithm again with two moments fixed, and so on until entanglement is detected or full tomography is achieved. In this setting, we need to distinguish the $k!$ different ordered arrangements of each $k$-element subset of $\mathcal{M}$. 

In the following, we call an ordered sequence of measurements a
\textit{path}, and we denote it by $\gamma$. To distinguish it from a
set, we denote it as a tuple with round parentheses, such as $(M_x,
M_y, M_{xz})$. A path of length $k$ can 
be alternatively seen as a list of $k$ sets  of increasing
size given by the restriction of the path to the first $k'$
observables with $1\leq k'\leq k$. For instance for $k=3$ the path
$(M_x, M_{xz}, M_y)$ can be seen 
as the list $\{M_x\}$, $\{M_{x}, M_{xz}\}$, and $\{M_x, M_y, M_{xz}\}$
(as usual we write sets in lexicographical order since the order within a set does not matter). 

Considering all $8!$ paths of length 8 would require an exceedingly
long computational time. For this reason, we slightly simplify the
problem by fixing the first measurement to perform. The most
reasonable choice, looking at the results in Fig.~\ref{prob}, is to
fix it as a diagonal observable $M_{xx}$, $M_{yy}$ or $M_{zz}$,
since for \textsc{$k=1$} it detects the largest fraction of entangled
states.  
{Up to relabelling of the axes, we can take $ M_{xx} $ as first
  element, since, as before, we only keep non-equivalent
  paths. To find these paths, we define a canonical representation of a path
  $\gamma$ of length $k$ by considering its equivalent list of $k$
  sets of length $k'$. For each of these sets we choose the first 
  one in lexicographical order among the ones that are obtained by relabelling of the
  axes. The list of $k$ sets obtained in this way is the canonical
  representation of $\gamma$.   
Two paths are equivalent if they have the same
canonical representation. We report the number $m_k'$ of
non-equivalent paths of length $k$ in Table \ref{t1}, where \eg for $
k=2 $ the non-equivalent sequences will be $(M_{xx},M_x)$
,$(M_{xx},M_y)$, $(M_{xx},M_{xy})$, $(M_{xx},M_{yy})$, and
$(M_{xx},M_{yz})$.}

\subsection{Path probabilities}
We now show how to retrieve the results for this more general case from the $ p_I^{(k)} $ obtained in the previous section.
The probability to detect the state as entangled after the first measurement, say $M_1$, is $P(E,\lbrace M_1\rbrace)$, given by the previous section. The probability to detect the state as entangled after the second measurement, say $M_2$, is then $ P(E,\lbrace M_1,M_2 \rbrace| \bar{E},\lbrace M_1 \rbrace)$, which is  the probability of detecting entanglement with the second measurement given that it was not detected with the first one. 
This quantity now depends on which measurement is performed first. This is illustrated in Fig.~\ref{sketch}.
\begin{figure}[htpb]
\includegraphics[scale=0.3]{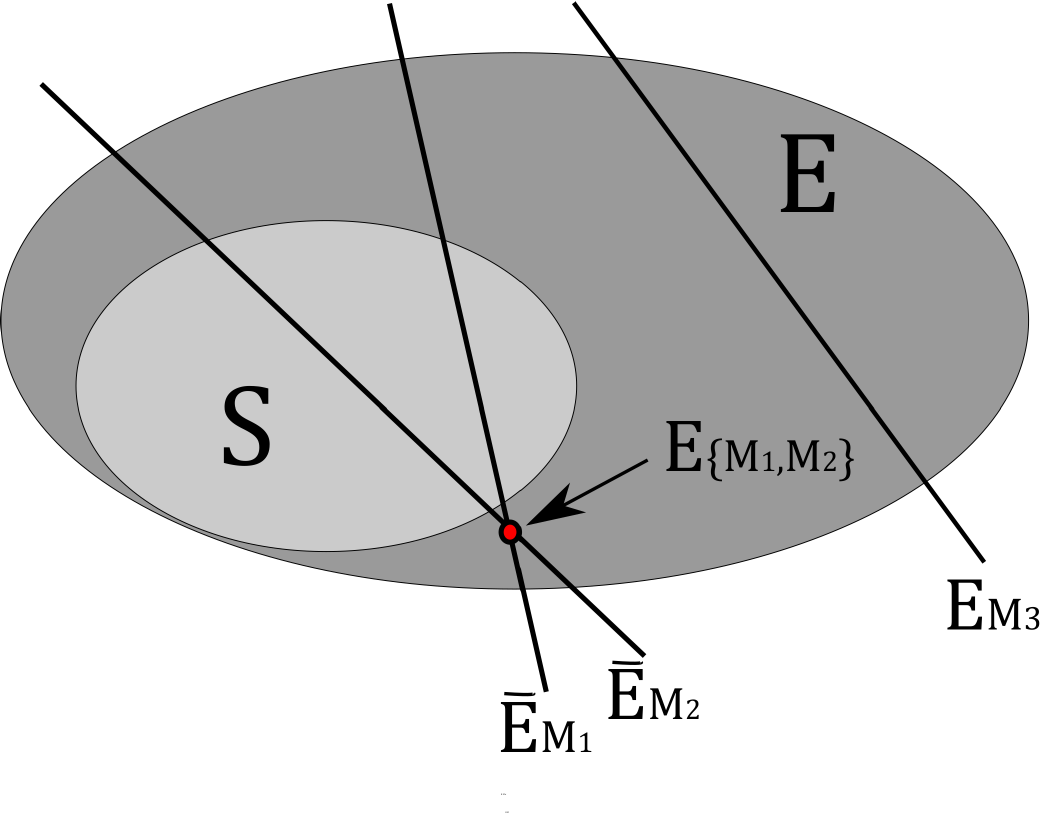}
\caption{{Two-dimensional sketch of the sets involved. $ S\equiv$ separable states  and $ E\equiv$ entangled states; we consider an arbitrary state in region $E$. Fixing one moment means restricting the set of compatible states to a hyperplane (one of the three lines in the sketch). Hyperplanes which cross the set of separable states contain both entangled and separable states, thus measuring observables $M_1$ or $M_2$ alone is not enough to detect entanglement. Fixing both on the other hand restricts the set of compatible states to a region (a point in the sketch) outside S, \ie observables $\{M_1,M_2\}$ together detect a fraction of states as entangled (\textit{E}). The third line instead does not cross the set of separable states, meaning that measuring $M_3$ suffices to detect entanglement (which we denote $ E_{M_3} $}).}
\label{sketch}
\end{figure}
Using the theorem of total probability, we have
\begin{align}
P(E, \lbrace M_1, M_2 \rbrace)=& P(E,\{M_1\})P(E, \lbrace M_1, M_2 \rbrace \vert E,\{M_1\})+ \nonumber \\
& P(\bar{E},\{M_1\})P(E, \lbrace M_1, M_2 \rbrace \vert \bar{E},\{M_1\})\;.
\label{thm}
\end{align}
Since  $ P(\bar{E},\{M_1\})=1-P(E,\{M_1\}) $ and $P(E,\lbrace M_1, M_2 \rbrace \vert E,M_1)=1$ we get
\begin{equation}
\label{pcond}
P(E, \lbrace M_1, M_2 \rbrace \vert \bar{E},\{M_1\})=\frac{P(E, \lbrace M_1,M_2 \rbrace)-P(E,\{M_1\})}{1-P(E,\{M_1\})}\;.
\end{equation}
Thus, the conditional probability we are looking for can be expressed solely in terms of the $p_I^{(k)}$ of the previous section. 

Let then $ \gamma=(M_1,...,M_8)$ be a path of length $k=8$. We define 
\begin{equation}
q^{(k)}(\gamma)=P(E,\lbrace M_1,...,M_k\rbrace\vert \bar{E}, \lbrace M_1,...,M_{k-1}\rbrace)
\end{equation}
as the probability of detecting entanglement at step \textit{k} in $ \gamma $ given that no entanglement was detected up to the step $ k-1 $. By a reasoning similar to the one leading to Eq.~\eqref{pcond}, we can express $ q^{(k)}(\gamma) $ in terms of the $ p^{(k)}(\gamma)\equiv P(E,\lbrace M_1,...,M_k\rbrace) $, as
\begin{equation}
\label{qkgamma}
q^{(k)}(\gamma)=\frac{p^{(k)}(\gamma)-p^{(k-1)}(\gamma)}{1-p^{(k-1)}(\gamma)}\;.
\end{equation}
In particular, since $p^{(0)}=0$ (as nothing is measured, and hence detected as entangled, at level 0), we have $q^{(1)}=p^{(1)}$. 
Inverting \eqref{qkgamma} one obtains $ p^{(k)}(\gamma) $ in terms of $ q^{(k)}(\gamma)$ as $ p^{(k)}(\gamma)=\sum_{j=1}^k q^{(j)}(\gamma) \prod_{n=j+1}^k (1-q^{(n)}(\gamma))$.

A third natural probability to consider is related to our measurement
algorithm, where we perform TMS calculations at step $k$ only if the
state was compatible with a separable state. We define $
r^{(k)}(\gamma) $ as the probability of stopping exactly at the $k$th
level when measurements are taken along path $\gamma$.  It can be
written as the joint probability $ P(E, \lbrace M_1,...,M_k \rbrace
\cap \bar{E},\lbrace M_1,...,M_{k-1}\rbrace) $. Using the identity
$P(A\cap B)=P(A|B)P(B)$, $r^{(k)}(\gamma)$ can be expressed as $
q^{(k)}(\gamma)(1-p^{(k-1)}(\gamma)) $. It can be rewritten in
terms of $ q^{(k)}(\gamma) $ or $ p^{(k)}(\gamma) $ as 
\begin{equation}
r^{(k)}(\gamma)=q^{(k)}(\gamma)\prod_{j=1}^{k-1}(1-q^{(j)}(\gamma))=p^{(k)}(\gamma)-p^{(k-1)}(\gamma)\;.
\label{rk}
\end{equation}

\subsection{Best path}
Using \eqref{qkgamma}--\eqref{rk} and the numerical results of the previous
section, we can obtain a numerical estimate of the $q^{(k)}(\gamma)$
and the $r^{(k)}(\gamma)$
for all possible paths. The optimal path $ \gbest$ is the one that detects as
quickly as possible (on average) if the state is entangled. To
identify $\gbest$ among all possible ones we define the
average depth at which our algorithm 
stops as
\begin{equation}
d(\gamma)=\sum_{k=1}^8 k r^{(k)}(\gamma)\;.
\label{dpath}
\end{equation}
Expressing Eq.~\eqref{dpath} in words, $ d(\gamma) $ gives the 
number of measurements that one needs to perform, 
on 
average, to detect a state as entangled, following the path $\gamma$.
Each path will be characterized by this number and in particular the
shortest path will be given by  
\begin{align}
\gbest &= \underset{\gamma\in S}{\text{arg 
min}} \hspace{0.1cm} d(\gamma) \,. 
\end{align}
The distribution of $ d(\gamma) $ 
over all 3228 paths of length eight for symmetric states of two qubits
is reported in Fig.~\ref{hist}.
\begin{figure}[t]
\includegraphics[scale=0.37]{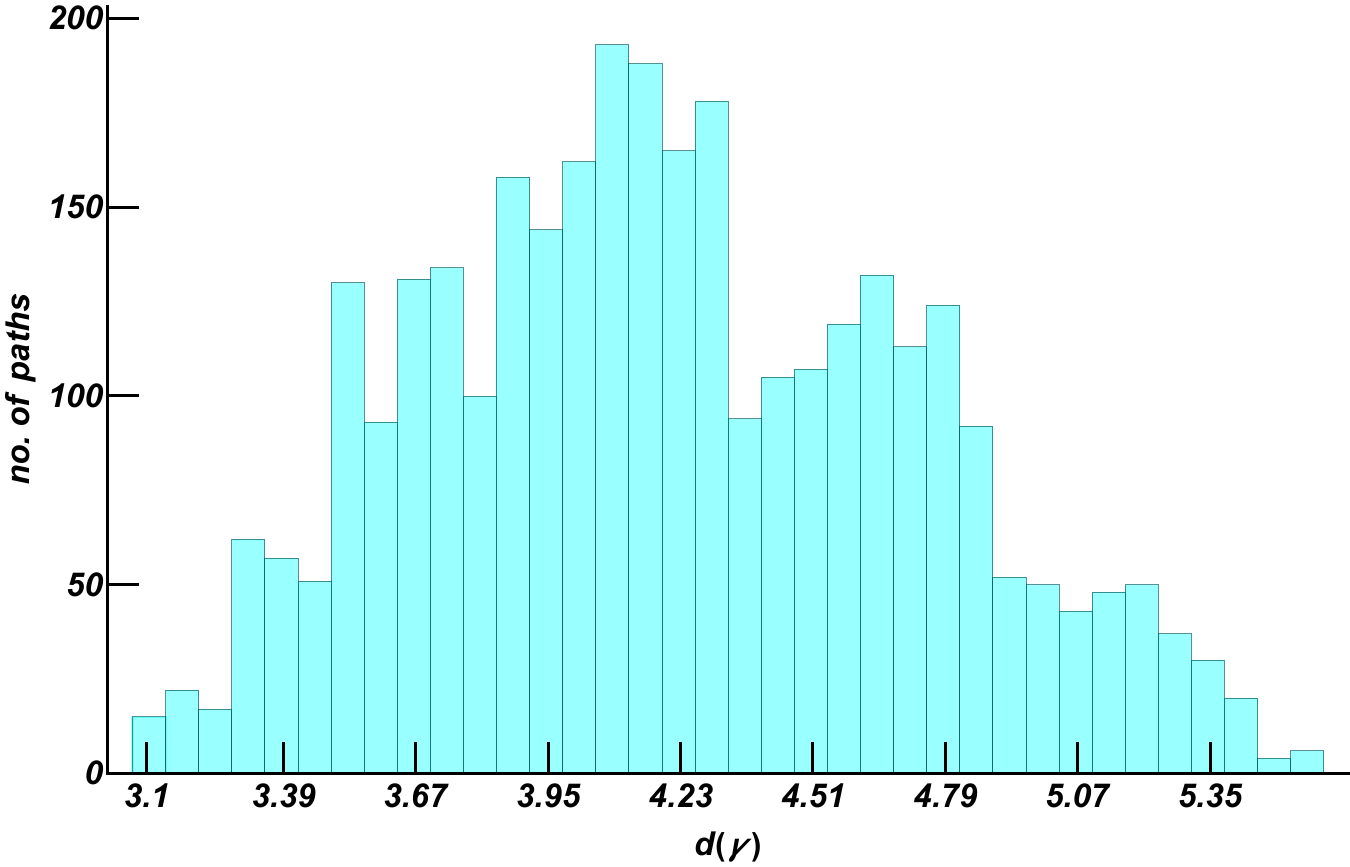}
\caption{Distribution (with bin width $ 0.07 $) of lengths $ d(\gamma) $ 
of measurement
  sequences $\gamma$ of symmetric states of two qubits 
resulting in detection of entanglement between the minimum value of $
3.07 $ and the maximum one of $ 5.61 $.} 
\label{hist}
\end{figure}  
The minimum value found for $ d(\gamma) $ is $ d=3.07 $, while the
maximum value is $5.61$. The minimum value is degenerate and
corresponds to three optimal paths. Although these three paths do not have the
same canonical representation they lead to the same value because of 
condition \eqref{traceless}.  If
one considered that knowing two diagonal moments is equivalent to
knowing them all, and included in the symmetrization the third diagonal
moment once the first two are measured, there would be a unique
optimal path. We report here one of
the three equivalent optimal paths: $\gbest=(M_{xx},M_{yy},
M_{xz},M_{yz}, M_{xy},M_{x}, M_{y},M_{z})$; choosing this path, one
only needs to perform (in average) three measurements to detect a
state as entangled. These three measurements give access to the two
diagonal moments (and thus all of them via \eqref{traceless}), and one
of the off-diagonal ones. 
The probabilities relative to this best path are shown in Fig.~\ref{pbest}.
\begin{figure}[htpb]
\includegraphics[scale=0.27]{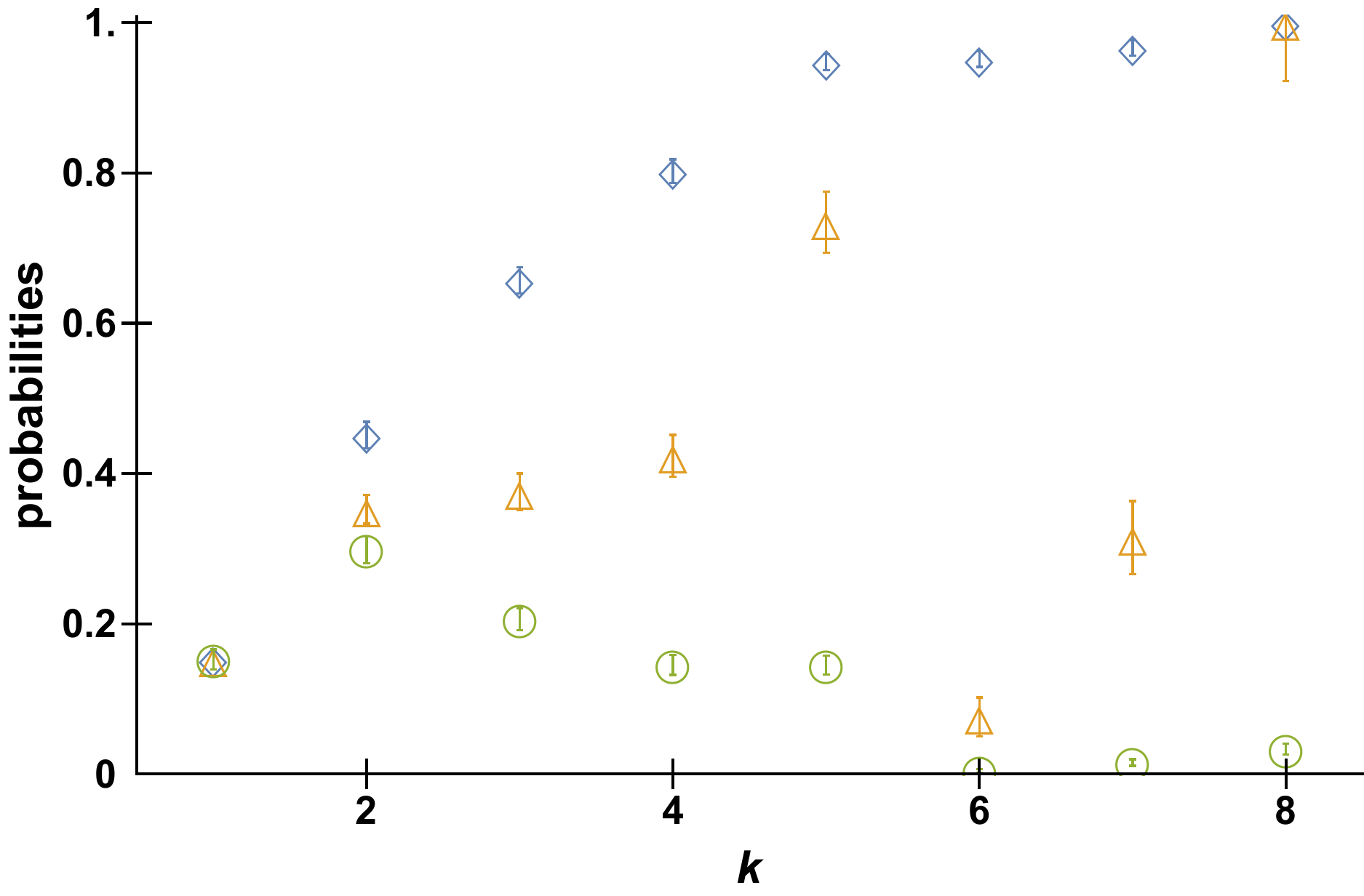}
\caption{ Probabilities $ p^{(k)}(\gbest) $ 
(blue diamonds), $ q^{(k)}(\gbest) $ (orange triangles), and $
r^{(k)}(\gbest) $ (green 
circles). The 
error bars represent the statistical 
errors 
 derived from those of the $p_I^{(k)}$, see Fig.~\ref{prob}.} 
\label{pbest}
\end{figure}

Rewriting $ d(\gamma) $ in terms of $p^{(k)}(\gamma)$ we get
\begin{equation}
d(\gamma)=8 p^{(8)}(\gamma)-p^{(7)}(\gamma)-p^{(6)}(\gamma)-...-p^{(1)}(\gamma)
\end{equation}
It turns out that choosing measurements according to $\gbest$
coincides (within the error bars) with choosing for each $k$, $ 1 \leq
k \leq 8 $, the best set of measurements, \ie the one with the highest
probability of detecting entanglement at a given level (highest $
p_I^{(k)} $ among the $ m_k $ possibilities for each \textit{k}). This
is not obvious, and it is not always the case: 
a counter-example is given by a binary tree of depth 4 in which the random
probabilities satisfy the same constraint as in our case, \ie $
p^{(k-1)}(\gamma)\leq p^{(k)}(\gamma)$ and the four paths have
probabilities: $
(0.57,0.62,0.76),(0.57,0.62,0.95),(0.57,0.68,0.77),(0.57,\\0.68,0.78)
$. It is easily verified that the best path, with $ d(\gamma)=1.8 $,
is the second one, which at depth $2$ does not have the highest $
p^{(2)}(\gamma) $, so the minimal $ d(\gamma)$ does not always
correspond to the path with the highest $ p^{(k)}(\gamma)$ at each
step.\\
In practice, joint measurements such as $M_{xx}$ might be more
challenging to implement than two single measurements $M_x$, as qubits
need a unitary operation to entangle them first and then two local
measurements. In such a case, one might modify \eqref{dpath} with
another factor for each path that takes such additional costs into
account. Also, we base our analysis on average values of measurement
outcomes which we took as known with arbitrary precision.  This is, of
course, an idealization.  In practice, only a finite number of
measurements can be performed, leading to statistical error bars for
each moment.  These can in principle be taken into account in the TMS
algorithm, but increase computational time. 
Both of these points are beyond the scope of the present paper.

\section{Non-symmetric case}
So far we restricted ourselves to symmetric states of two qubits. Let us now consider the generic case of arbitrary two-qubit non-symmetric states. In this case we can still exploit the TMS algorithm, with the following differences \cite{tms}. The bipartite state acts on the tensor product $ \mathcal{H}_1\otimes\mathcal{H}_2 $ of Hilbert spaces and each of them has now its own set of variables $x,y$ and $z$; we will label these variables as ($ x_i,y_i,z_i $), with $ i=1,2 $. The compact set \textit{K} is now the product of two Bloch spheres. The set $ \mathcal{M} $ of possible measurements is
\begin{align}
\mathcal{M}=\lbrace & M_{x_1},M_{y_1},M_{z_1},M_{x_2},M_{x_1x_2},M_{y_1x_2},M_{z_1x_2}, M_{y_2},\nonumber\\ 
& M_{x_1y_2},M_{y_1y_2},M_{z_1y_2},M_{z_2},M_{x_1z_2},M_{y_1z_2},M_{z_1z_2}\rbrace\;.
\end{align}
For example, $ M_{x_1} $ is the measurement of $ \sigma_x \otimes
\mathbb{1} $ and $ M_{x_1x_2} $ is the measurement of the joint
operator $ \sigma_x \otimes \sigma_x $. Up to relabelling the variables
for each qubit, some sets of measurement operators are equivalent. The
number $ m_k $ of non-equivalent sets of measurements is obtained by
applying the 36 possible permutations on the ($ x_i,y_i,z_i $). This
number is reported in Table \ref{t2} for $ 1\leq k \leq 15 $. 
 \begin{table}[htpb]
  \begin{tabular}{ c|c|c|c|c|c|c|c|c|c|c|c|c|c|c|c }
    \hline
    $k$ & 1 & 2 & 3 & 4 & 5 & 6 & 7 & 8 & 9 & 10 & 11 & 12 & 13 & 14 & 15\\ \hline
    $m_k$ & 3 & 10 & 30 & 69 & 132 & 205 & 254 & 254 & 205 & 132 & 69 & 30 & 10 & 3 & 1 \\ \hline
  \end{tabular}
  \caption{Number $ m_k $ of non-equivalent unordered sets of
    measurements 
of two qubits 
for $ 1\leq k \leq 15 $.}
  \label{t2}
 \end{table}
 
The number $ m_k $ increases fast with $k$, and so does the size of
the moment matrices considered in the TMS algorithm: indeed, because
of condition \eqref{rank}, the algorithm always searches at least for
the first extension; in both cases (symmetric and non-symmetric) the
smallest extension corresponds to the moment matrix of order $ 2 $. In
the symmetric case it is a $ 10\times 10 $ matrix, while in the non
symmetric case it already becomes a $ 28 \times 28 $ matrix which
contains all the monomials up to degree 4 for the set of $6$ variables
$ x_i,y_i,z_i $, with $ i=1,2 $, \ie $210$ moments versus $ 35 $ in
the symmetric case. For the previous reasons computational times
become an issue in the non-symmetric case. Nevertheless we could estimate probabilities up to $k=5$, running the TMS algorithm over a database of 50000 non-symmetric two-qubit random states.
What we observe is that no state is detected as entangled with only one measurement, a tiny fraction ( $\sim 1 \% $) is detected as entangled by the combination of two measurements $ \lbrace M_{x_1 x_2},M_{y_1 y_2}\rbrace $, and the biggest fraction of states detected as entangled for $ 3\leq k\leq 5 $ are given respectively by the set of measurements $ \lbrace M_{x_1 x_2},M_{y_1 y_2},M_{z_1 z_2}\rbrace $ ($ \sim 10 \%$), $ \lbrace M_{x_1 x_2},M_{x_1 y_2},M_{y_1 x_2},M_{z_1 z_2}\rbrace $ ($ \sim 12 \%$), $\lbrace M_{x_1 x_2},M_{x_1 y_2},M_{y_1 x_2},M_{y_1 y_2},M_{z_1 z_2}\rbrace$ ($ \sim 23 \%$). This is a big difference compared to the symmetric case, in which we could detect $ \sim 15\% $ of the states as entangled with a single measurement, $ \sim 40\% $ already with two measurements and almost all states with five measurements.

}

\section{Higher spin-j}
Going back to the case of symmetric states, we can also 
get 
an idea of how complexity changes for higher spin sizes; indeed, the size of the set $ \mathcal{M} $ in the symmetric case corresponds to the sum of the number of monomials in three variables up to degree $ d=2j+1 $, where $ j $ is the spin size. These numbers form the sequence of triangular numbers $ T_n=\sum_{i=1}^n i=\frac{n(n+1)}{2} $; we can then write that $ m_k $ for any spin-$ j $ is
\begin{equation}
m_k=\binom{\sum_{n=1}^{2j+1}T_n-1}{k}
\end{equation}
where we subtract $ 1 $ since the first element of $ \mathcal{M} $ is always the identity.
However, in this case, we can still have some information looking at
the expression for the tensor representation of a separable state in
\eqref{sep}. Indeed, for an even number of qubits (integer spins) we
can look at the diagonal tensor entries, which are defined as the
entries of the form $X_{\mu_1...\mu_j \mu_1... \mu_j}$ with $
0\leq\mu_i\leq 3 $. These correspond to terms of the form $\sum_j
\omega_j(n_{\mu_1}...n_{\mu_j})^{2j} $; it follows that for a
separable state these entries are positive, since the $n_{\mu_i}$ are
real and $\omega_j\geq 0$. Therefore measuring a negative value for
any of the corresponding measurement operators means detecting
entanglement; we indicate the operators corresponding to the diagonal
entries of the tensor $ X_{\mu_1\mu_2...\mu_N} $ with $ \lbrace
D\rbrace_I $. 
We can then restrict our investigation for an integer spin $ j $ to these $ 4^j $ observables, which are further reduced by symmetry to $ {j+3}\choose{3} $.
We report in Fig. \ref{diagtot} the number of entangled states that
are not detected by any of the observables $ \lbrace D\rbrace_I $ for
spin size $ 1\leq j \leq  5 $ (for each $ j $ we used a sample of $
10^6 $ random states from which we again removed the separable
ones). The number of not detected entangled states decreases with the
spin size $ j $ and already for $j=4$ all the states in the sample are
detected; we can also observe that restricting the analysis to these
observables already gives  significant information for spin-$ 1 $ and
spin-$ 2 $ and almost complete information for spin-$ 3 $. 
\begin{figure}
\includegraphics[scale=0.8]{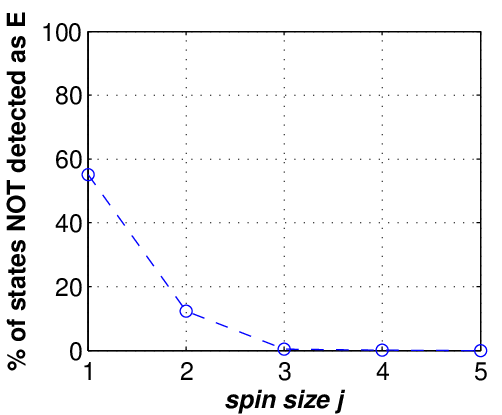}
\caption{
Percentage of entangled states not detected by any of the negative outcomes of the measurements $ \lbrace D\rbrace_I $ corresponding to the diagonal entries of the tensor $ X_{\mu_1\mu_2...\mu_N} $ as a function of the spin size $ j $.}
\label{diagtot}
\end{figure}
Moreover, we can also compare these observables to see which is the
most efficient measurement to perform as we did for the spin-$1 $
case; to estimate the corresponding $ p_I^{(1)} $, we will again only
consider sets which are non-equivalent under permutation of the axes,
performing the transformations $ \lbrace P_{xy}, P_{xz}, P_{yz},
P_{yzx},P_{zxy} \rbrace$ described in section \ref{symandmeas}. The
results are shown in Fig. \ref{diagcomp}. 
\begin{figure}
\includegraphics[scale=0.8]{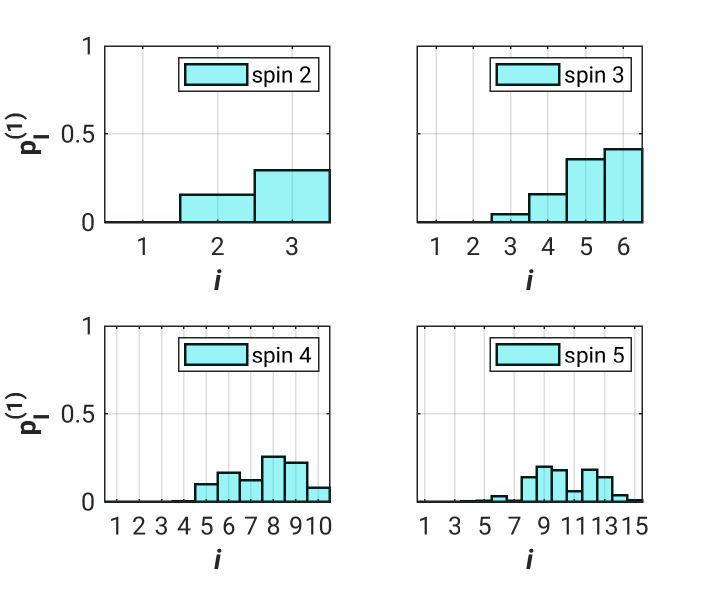}
\caption{Comparison of the non-equivalent 
diagonal 
observables $ \lbrace D\rbrace_I $ for spin-$ j $, $ 2\leq j \leq 5
$: Probabilities $p^{(1)}_I$ as a function of the label $i$ of the set $I$. The highest values 
are reached, respectively, for 
$D_{xxyy},D_{xxyyzz}, D_{xxxxxxyy}, D_{xxxxyyyy}$ (where the last term
corresponds to the measurement of $
\mathbb{1}^{\otimes 2}\otimes\sigma_x^{\otimes
  4}\otimes\sigma_y^{\otimes 4} $; 
see Appendix \ref{app.B} for the
full list). 
}
\label{diagcomp}
\end{figure}
The question arises whether
similarly efficient measurements can be found 
for half-integer spin $ j $. It was recently shown in \cite{ppt} 
how the positive-partial-transpose (PPT) separability criterion for
symmetric states of multi-qubit systems can be formulated in terms of
matrix inequalities based on the tensor representation in
Eq.~\eqref{tensrepr}. It is possible to construct a matrix $ T $ from
the tensor representation of the state and show that it is similar to
the partial transpose of the density matrix written in the
computational basis. In the case of spin-$ 3/2 $ this matrix is a $
8\times 8 $ Hermitian matrix given by $ T_{\mu i,\nu
  i'}=\sum_{\tau=0}^3 X_{\tau\mu\nu}\sigma^{\tau}_{i,i'} $, where $
\sigma^{\tau}_{i,i'} $ are the Pauli-matrix components, and its
positivity is a necessary and 
sufficient classicality criterion; as a consequence, the positivity of
the diagonal entries is a necessary condition for a separable
state. We can again restrict our investigation to the corresponding
observables $ \lbrace D\rbrace_I $, but this time it implies the
measurement of sets of two observables. Indeed, in terms of the tensor
entries $ X_{\mu_1\mu_2\mu_3} $, the diagonal entries of $ T $ are
$X_{000}\pm X_{003}, X_{011}\pm X_{113}, X_{022}\pm X_{223},X_{033}\pm
X_{333} $, so we need to compare pairs of outcomes. Recalling that $
X_{000}=1 $, we can neglect the first entry, since the condition $ -1
\leq X_{003} \leq 1  $ is always satisfied. The results of such
investigation for the other six pairs and for their combinations (all the $ {6}\choose{k} $ sets, with $ 2\leq k\leq 6 $ ) are reported in Fig. \ref{diag32}. As before, we can gain already relevant information from this restricted analysis. \\
\begin{figure}
\includegraphics[scale=0.8]{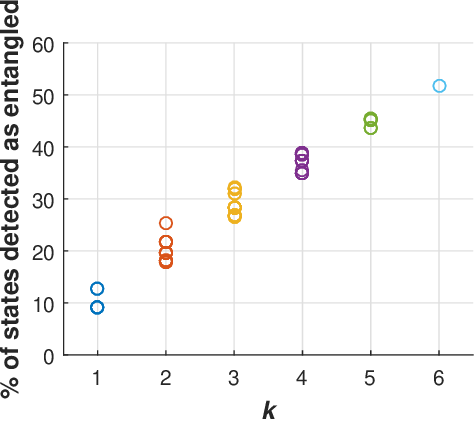}
\caption{
Entanglement detection probabilities based on the
  negativity of the
$\binom{6}{k}$ subsets of the set $\{
  X_{011}-X_{113},X_{011}+X_{113}, X_{022}-X_{223},  X_{022}+X_{223},
  X_{033}-X_{333}, X_{033}+X_{333} \}$ for $k=1,\ldots,6$, where the tensor $  
  X_{\mu_1\mu_2\mu_3} $ represents a spin-$ 3/2 $.} 
\label{diag32}
\end{figure}
\\

\section{Conclusions}
In summary, we have studied the statistics of lengths of measurement
sequences for multi-qubit systems that allow one to detect
entanglement without any prior information about the state, both for
unordered sets of measurements and ordered ones (i.e.~measurement
paths).   For symmetric 
states of two qubits, we have identified the best measurement
path that results, on average over all randomly chosen entangled
states, in a proof of entanglement with 3.07 measurements (compared to
8 measurements needed for full tomography in this case). For larger
numbers $N$ of qubits in symmetric states, we found that measurements
based on the diagonal matrix elements of the moment matrix of the
state become very efficient in detecting entanglement.  Their number
grows like $N^3$, and already for $N=8$ qubits the number of states
{\em not} detected as entangled has decayed to about $10^{-6}$ or
smaller. For non-symmetric states, substantially larger numbers of
measurements are needed to detect entanglement with certainty: at
least two measurements are 
needed for two-qubit states, resulting in only about 1\% detection
probability, however.  With five measurements the probability increases
to about 23\%.  
The work is based on the
truncated moment sequence 
algorithm that naturally allows one to deal with missing data.  It is
very flexible and can be easily adapted to experimentally relevant
ensembles of states and other side-conditions, such as sets of 
measurements that can be implemented, or more elaborate cost
functions. 

\section*{Acknowledgments} DB thanks OG, the LPTMS, and the 
Universit\'e Paris-Saclay for hospitality.

\appendix

\section{Unordered measurement sets}
\label{app}
We list here all the $ m_k $ unique sets of $ k $ measurements for $ 1\leq k\leq 8 $.
\begin{table}[H]
  \begin{tabular}{c|c}
    \hline
    $k=1$ \\ \hline 1 & $M_x$ \\2 & $M_{xx}$ \\3 & $M_{xy}$ \\ \hline
  \end{tabular}
  \begin{tabular}{c|c}
    \hline
    $k=2$ \\ \hline 1 & $\lbrace M_x,M_y\rbrace $ \\2 & $\lbrace M_x,M_{xx}\rbrace $ \\3 & $\lbrace M_x,M_{xy}\rbrace $ \\4 & $\lbrace M_x,M_{yy}\rbrace $ \\5 & $\lbrace M_x,M_{yz}\rbrace $ \\6 & $\lbrace M_{xx},M_{xy}\rbrace $ \\7 & $\lbrace M_{xx},M_{yy}\rbrace $ \\8 & $\lbrace M_{xx},M_{yz}\rbrace $ \\9 & $\lbrace M_{xy},M_{xz}\rbrace $ \\ \hline
  \end{tabular}
  \begin{tabular}{c|c}
    \hline
    $k=3$ \\ \hline 1 & $\lbrace M_x, M_y, M_z\rbrace $ \\2 & $\lbrace M_{x}, M_{y}, M_{xx}\rbrace $\\3 & $\lbrace M_{x}, M_{y}, M_{xy}\rbrace $\\4 &  $\lbrace M_{x}, M_{y}, M_{xz}\rbrace $ \\5 & $\lbrace M_{x}, M_{z}, M_{yy}\rbrace $ \\6 & $\lbrace M_{x}, M_{xx}, M_{xy}\rbrace $ \\7 & $\lbrace M_{x}, M_{xx}, M_{yy}\rbrace $\\8 & $\lbrace M_{x}, M_{xx}, M_{yz}\rbrace $ \\9 & $\lbrace M_{x}, M_{xy}, M_{xz}\rbrace $ \\10 & $\lbrace M_{x}, M_{xy}, M_{yy}\rbrace $ \\11 & $\lbrace M_{x}, M_{xy}, M_{yz}\rbrace $\\12 & $\lbrace M_{x}, M_{xz}, M_{yy}\rbrace $ \\13 & $\lbrace M_{x}, M_{yy}, M_{yz}\rbrace $ \\14 & $\lbrace M_{z}, M_{xx}, M_{yy}\rbrace $\\15 & $\lbrace M_{xx}, M_{xy}, M_{xz}\rbrace $ \\16 & $\lbrace M_{xx}, M_{xy}, M_{yy}\rbrace $ \\17 & $\lbrace M_{xx}, M_{xy}, M_{yz}\rbrace $ \\18 & $\lbrace M_{xx}, M_{xz}, M_{yy}\rbrace $ \\19 & $\lbrace M_{xy}, M_{xz}, M_{yz}\rbrace$
    \end{tabular}
    \end{table}
    
    \begin{table}[H]
       \centering
    \begin{tabular}{c|c}
    \hline
    $k=4$ \\ \hline
1 & $\lbrace M_{x}, M_{y}, M_z, M_{xx} \rbrace $ \\2 & $\lbrace M_{x}, M_{y}, M_{z}, M_{xy} \rbrace $ \\3 & $\lbrace M_{x}, M_{y}, M_{xx}, M_{xy} \rbrace $ \\4 & $\lbrace M_{x}, M_{y}, M_{xx}, M_{xz} \rbrace $ \\5 & $\lbrace M_{x}, M_{y}, M_{xx}, M_{yy} \rbrace $ \\6 & $\lbrace M_{x}, M_{y}, M_{xx}, M_{yz} \rbrace $  \\7 & $\lbrace M_{x}, M_{z}, M_{xx}, M_{yy} \rbrace $ \\8 & $\lbrace M_{x}, M_{y}, M_{xy}, M_{xz} \rbrace $  \\9 & $\lbrace M_{x}, M_{z}, M_{xz}, M_{yy} \rbrace $ \\10 & $\lbrace M_{x}, M_{y}, M_{xz}, M_{yz} \rbrace $ \\11 & $\lbrace M_{x}, M_{z}, M_{xy}, M_{yy} \rbrace $ \\12 & $\lbrace M_{x}, M_{xx}, M_{xy}, M_{xz} \rbrace $ \\13 & $\lbrace M_{x}, M_{xx}, M_{xy}, M_{yy} \rbrace $ \\14 & $\lbrace M_{x}, M_{xx}, M_{xy}, M_{yz} \rbrace $ \\15 & $\lbrace M_{x}, M_{xx}, M_{xz}, M_{yy} \rbrace $ \\16 & $\lbrace M_{x}, M_{xx}, M_{yy}, M_{yz} \rbrace $ \\17 & $\lbrace M_{x}, M_{xy}, M_{xz}, M_{yy} \rbrace $ \\18 & $\lbrace M_{x}, M_{xy}, M_{xz}, M_{yz} \rbrace $ \\19 & $\lbrace M_{x}, M_{xy}, M_{yy}, M_{yz} \rbrace $ \\20 & $\lbrace M_{z}, M_{xx}, M_{yy}, M_{yz} \rbrace $ \\21 & $\lbrace M_{x}, M_{xz}, M_{yy}, M_{yz} \rbrace $ \\22 & $\lbrace M_{z}, M_{xx}, M_{xz}, M_{yy} \rbrace $ \\23 & $\lbrace M_{xx}, M_{xy}, M_{xz}, M_{yy} \rbrace $ \\24 & $\lbrace M_{xx}, M_{xy}, M_{xz}, M_{yz} \rbrace $ \\25 & $\lbrace M_{xx}, M_{xz}, M_{yy}, M_{yz} \rbrace $ 
\end{tabular}
\end{table}

\begin{table}[H]
\centering
\begin{tabular}{c|c}
    \hline
    $k=5$ \\ \hline
1 & $\lbrace M_{x}, M_y, M_z , M_{xx}, M_{xy}\rbrace $ \\2 & $\lbrace M_{x}, M_{y}, M_{z}, M_{xx}, M_{yy} \rbrace $ \\3 & $\lbrace M_{x}, M_{y}, M_{z}, M_{xx}, M_{yz} \rbrace $ \\4 & $\lbrace M_{x}, M_{y}, M_{z}, M_{xy}, M_{xz} \rbrace $ \\5 & $\lbrace M_{x}, M_{y}, M_{xx}, M_{xy}, M_{xz} \rbrace $  \\6 & $\lbrace M_{x}, M_{y}, M_{xx}, M_{xy}, M_{yy} \rbrace $ \\7 & $\lbrace M_{x}, M_{y}, M_{xx}, M_{xy}, M_{yz} \rbrace $  \\8 & $\lbrace M_{x}, M_{z}, M_{xx}, M_{xz}, M_{yy} \rbrace $ \\9 & $\lbrace M_{x}, M_{y}, M_{xx}, M_{xz}, M_{yy} \rbrace $ \\10 & $\lbrace M_{x}, M_{y}, M_{xx}, M_{xz}, M_{yz} \rbrace $ \\11 & $\lbrace M_{x}, M_{z}, M_{xx}, M_{xy}, M_{yy} \rbrace $ \\12 & $\lbrace M_{x}, M_{z}, M_{xx}, M_{yy}, M_{yz} \rbrace $ \\13 & $\lbrace M_{x}, M_{y}, M_{xy}, M_{xz}, M_{yz} \rbrace $ \\14 & $\lbrace M_{x}, M_{z}, M_{xy}, M_{xz}, M_{yy} \rbrace $ \\15 & $\lbrace M_{x}, M_{z}, M_{xy}, M_{yy}, M_{yz} \rbrace $ \\16 & $\lbrace M_{x}, M_{xx}, M_{xy}, M_{xz}, M_{yy} \rbrace $ \\17 & $\lbrace M_{x}, M_{xx}, M_{xy}, M_{xz}, M_{yz} \rbrace $\\18 & $\lbrace M_{x}, M_{xx}, M_{xy}, M_{yy}, M_{yz} \rbrace $ \\19 & $\lbrace M_{x}, M_{xx}, M_{xz}, M_{yy} , M_{yz}\rbrace $ \\20 & $\lbrace M_{x}, M_{xy}, M_{xz}, M_{yy} , M_{yz}\rbrace $ \\21 & $\lbrace M_{z}, M_{xx}, M_{xz}, M_{yy}, M_{yz} \rbrace $ \\22 & $\lbrace M_{z}, M_{xx}, M_{xy}, M_{yy}, M_{yz} \rbrace $ \\23 & $\lbrace M_{xx}, M_{xy}, M_{xz}, M_{yy}, M_{yz} \rbrace $ 
\end{tabular}
\end{table}

\begin{table}[H]
\centering
\begin{tabular}{c|c}
\hline
    $k=6$ \\ \hline
1 & $\lbrace M_{x}, M_{y},M_z, M_{xx}, M_{xy}, M_{xz} \rbrace $ \\ 
2 & $\lbrace M_{x}, M_y, M_z, M_{xx}, M_{xy},M_{yy} \rbrace $ \\ 
3 & $\lbrace M_{x}, M_y, M_z, M_{xx}, M_{xy},M_{yz} \rbrace $ \\
4 & $\lbrace M_{x}, M_{y}, M_{z}, M_{xx}, M_{xz},M_{yy} \rbrace $ \\ 
5 & $\lbrace M_{x}, M_y, M_z, M_{xy}, M_{xz},M_{yz} \rbrace $ \\
6 & $\lbrace M_{x},M_{y}, M_{xx}, M_{xy}, M_{xz}, M_{yy} \rbrace $ \\
7 & $\lbrace M_{x},M_{y}, M_{xx}, M_{xy}, M_{xz}, M_{yz} \rbrace $ \\ 
8 & $\lbrace M_{x}, M_{z}, M_{xx}, M_{xy},M_{xz}, M_{yy} \rbrace $ \\ 
9 & $\lbrace M_{x}, M_{z}, M_{xx}, M_{xz},M_{yy}, M_{yz} \rbrace $ \\ 
10 & $\lbrace M_{x}, M_y, M_{xx}, M_{xz},M_{yy}, M_{yz} \rbrace $ \\ 
11 & $\lbrace M_{x}, M_{z}, M_{xx}, M_{xy},M_{yy}, M_{yz} \rbrace $ \\ 
12 & $\lbrace M_{x}, M_{z}, M_{xy}, M_{xz},M_{yy}, M_{yz} \rbrace $ \\ 
13 &  $\lbrace M_{x}, M_{xx}, M_{xy}, M_{xz},M_{yy}, M_{yz} \rbrace $ \\ 
14 & $\lbrace M_{z}, M_{xx}, M_{xy}, M_{xz},M_{yy}, M_{yz} \rbrace $ 
\end{tabular}
\end{table}

\begin{table}[H]
\centering
\begin{tabular}{c|c}
\hline
    $k=7$ \\ \hline
1 & $\lbrace M_{x},M_{y},M_{z}, M_{xx}, M_{xy}, M_{xz}, M_{yy} \rbrace $ \\
2 & $\lbrace M_{x},M_{y},M_{z}, M_{xx}, M_{xy}, M_{xz}, M_{yz} \rbrace $ \\
3 & $\lbrace M_{x}, M_{y}, M_{z}, M_{xx},M_{xz}, M_{yy} ,M_{yz}\rbrace $ \\
4 & $\lbrace M_{x},M_{y}, M_{xx}, M_{xy}, M_{xz}, M_{yy},M_{yz} \rbrace $ \\
5 & $\lbrace M_{x}, M_{z}, M_{xx}, M_{xy},M_{xz}, M_{yy},M_{yz} \rbrace $ 
\end{tabular}
\end{table}

\begin{table}[H]
\centering
\begin{tabular}{c|c}
\hline
    $k=8$ \\ \hline
1 & $\lbrace M_x , M_{y} , M_{z} , M_{xx} , M_{xy} , M_{xz} , M_{yy} , M_{yz}\rbrace$ \\ 
\end{tabular}
\end{table}

\section{Non-equivalent diagonal observables}\label{app.B}

We list here all the non-equivalent observables $ D _I $ for spin-$j$, $ 2\leq j\leq 5 $.
\begin{table}[H]
  \begin{tabular}{c|c}
    \hline
    $j=2$ \\ \hline 1 & $D_{xx}$ \\2 & $D_{xxxx}$ \\3 & $D_{xxyy}$ \\ \hline
  \end{tabular}
  \begin{tabular}{c|c}
    \hline
    $j=3$ \\ \hline 1 & $  D_{xx}  $ \\2 & $  D_{xxxx}  $ \\3 & $  D_{xxyy}  $ \\4 & $  D_{xxxxxx}  $ \\5 & $  D_{xxxxyy}  $ \\6 & $  D_{xxyyzz}  $ \\ \hline
  \end{tabular}
  \begin{tabular}{c|c}
    \hline
    $j=4$ \\ \hline 1 & $  D_{xx}  $ \\2 & $  D_{xxxx}  $\\3 & $  D_{xxyy}  $\\4 &  $  D_{xxxxxx}  $ \\5 & $  D_{xxxxyy}  $ \\6 & $  D_{xxyyzz}  $ \\7 & $  D_{xxxxxxxx}  $\\8 & $  D_{xxxxxxyy}  $ \\9 & $  D_{xxxxyyyy}  $ \\10 & $  D_{xxxxyyzz}  $ \\ \hline
    \end{tabular}
    \end{table}
    
    \begin{table}[H]
    \centering
    \begin{tabular}{c|c}
    \hline
    $j=5$ \\ \hline
1 & $  D_{xx}  $ \\2 & $  D_{xxxx}  $\\3 & $  D_{xxyy}  $\\4 &  $  D_{xxxxxx}  $ \\5 & $  D_{xxxxyy}  $ \\6 & $  D_{xxyyzz}  $ \\7 & $  D_{xxxxxxxx}  $\\8 & $  D_{xxxxxxyy}  $ \\9 & $  D_{xxxxyyyy}  $ \\10 & $  D_{xxxxyyzz}  $ \\11 & $  D_{xxxxxxxxxx}  $ \\12 & $  D_{xxxxxxxxyy}  $ \\13 & $  D_{xxxxxxyyyy}  $ \\14 & $  D_{xxxxxxyyzz}  $ \\15 & $  D_{xxxxyyyyzz}  $ \\ \hline
\end{tabular}
\end{table}

\bibliographystyle{apsrev}
\bibliography{EntDetNM,mybibs_bt}

\end{document}